\documentclass[aps,prl,preprint,groupedaddress]{revtex4-2}
\usepackage[a4paper, total={6in, 9in}]{geometry}
\usepackage{xcolor}
\usepackage{xspace}
\usepackage{graphicx}
\usepackage{epstopdf, epsfig}
\usepackage{mathtools, amsmath, amssymb, amsfonts, bm}
\usepackage{cprotect}
\usepackage{algorithm, algpseudocode}
\usepackage{enumitem}
\usepackage{mathrsfs}
\usepackage{subcaption}
\usepackage{hyperref}
\usepackage{subcaption}

\definecolor{dgreen}{HTML}{008000}

\renewcommand{\vec}[1]{\textbf{#1}}

\makeatletter
\newcommand{\longdash}[1][2em]{%
\makebox[#1]{$\m@th\smash-\mkern-7mu\cleaders\hbox{$\mkern-2mu\smash-\mkern-2mu$}\hfill\mkern-7mu\smash-$}}
\newcommand{\bes}{\begin{enumerate}[wide, labelwidth=!, labelindent=0pt, label=\textbf{\textcolor{blue}{\arabic*}.}]}
\newcommand{\ees}{\end{enumerate}}
\newcounter{saveenumi}

\newcommand{\fig}[1]{Fig.~\ref{fig:#1}}

\newcommand{\uiso}{$U_\text{iso}$\xspace}

\newcommand{\sbm}{\ensuremath{\mathscr{S}}}
\newcommand{\sta}{\ensuremath{\sbm_{T_\alpha}}\xspace}

\graphicspath{{./figs/}}

\begin{document}

\title{A continuous symmetry breaking measure for finite clusters using Jensen-Shannon divergence}

\author{Ling Lan}
\affiliation{Department of Applied Physics and Applied Mathematics, Columbia University, New York, NY 10027, USA}

\author{Qiang Du}
\affiliation{Department of Applied Physics and Applied Mathematics, and Data Science Institute, Columbia University, NY 10027, USA}

\author{Simon J.~L.~Billinge}
\affiliation{Department of Applied Physics and Applied Mathematics, Columbia University, New York, NY 10027, USA\,}

\begin{abstract}
A quantitative measure of symmetry breaking is introduced that allows the quantification of which symmetries are most strongly broken due to the introduction of some kind of defect in a perfect structure.  The method uses a statistical approach based on the Jensen-Shannon divergence. 
The measure is calculated by comparing the transformed atomic density function with its original. Software code is presented that carries the calculations out numerically using Monte Carlo methods.
The behavior of this symmetry breaking measure is tested for various cases including finite size crystallites (where the surfaces break the crystallographic symmetry), atomic displacements from high symmetry positions, and collective motions of atoms due to rotations of rigid octahedra.
The approach provides a powerful tool for assessing local symmetry breaking and offers new insights that can help researchers understand  how different structural distortions affect different symmetry operations.
\end{abstract}

\maketitle

\section{Introduction}

Symmetry is a fundamental property in the analysis of systems ranging from particle physics to condensed matter materials, including crystals, molecules, and polymers, among others \cite{mainzerSymmetriesNatureHandbook1996, grossRoleSymmetryFundamental1996}. 
The role of symmetry breaking (SB) is equally significant, as it is closely associated with many critical phenomena in physics and material science, and has been extensively studied over the past decades \cite{arodzPatternsSymmetryBreaking2003, kibbleSpontaneousSymmetryBreaking2015}.
Traditionally, material symmetry has been viewed as a dichotomous concept, where a cluster of atoms either satisfies or violates a symmetry operation. 
Based on this understanding, symmetry finders have been developed to identify whether a given structure is invariant under certain transformations within a global tolerance~\cite{ivanovSimpleAlgorithmsDetermining1999, togoSpglibSoftwareLibrary2024}.

In nanomaterials, the structures tend to be complex and length-scale dependent. In general, we would like to explore the concept of distance-dependent point symmetry because local symmetry can differ from average symmetry, for example, due to the averaging of local symmetry broken distortions over multiple allowed variants~\cite{qiuOrbitalCorrelationsPseudocubic2005, bozinLocalSnDipolarCharacter2023, zhaoIntrinsicLocalSymmetry2022, xieHiddenLocalSymmetry2022}. 
In this case, some symmetries may be absolutely or approximately preserved by the averaging whereas other symmetries are significantly broken.
This raises new questions that are not addressed by the dichotomous view of material symmetry.
For example, one may inquire about the extent and manner in which material properties are altered when symmetries are only approximately broken. 
In this case, we may be able to ignore the weak breaking of some symmetries in our analysis of the material, but not ignore others that have a stronger effect on the properties. 
Additionally, it may be of interest to determine which symmetry operations are significantly or only approximately violated by a specific structural distortion. 
To address this issue, we aim to develop a quantitative, rather than categorical, measure of SB that bridges the gap between satisfying and violating symmetries. 

The concept of creating a continuous SB measure for material structures and molecular systems has long been a subject of interest. For example, Zabrodsky et al.~\cite{zabrodskyContinuousSymmetryMeasures1992}, and the software based on their method \cite{tuvi-aradCSMSoftwareContinuous2024}, define a continuous measure as the minimal distance between a given structure and a structure with the desired symmetry. They first scan all relevant permutations to find the reference structure with the desired symmetry and then calculate the measure as the squared error of the coordinates. 
Similar methods have been applied in other studies \cite{keinanContinuousSymmetryAnalysis2001, keinanStudiesCopperIIComplexes2001}. 
However, this approach has two main limitations: the number of permutations increases significantly for larger structures, and it only considers atomic coordinates, ignoring other particle properties. In contrast, our paper introduces a continuous SB measure from a statistical perspective. 
We avoid the search for the desired structure by adopting a continuous representation of finite clusters. 
And this measure not only accounts for the positions of particles but also incorporates atomic species, occupancy, and atomic distributions through thermal vibrations.

Our statistical symmetry breaking measure (SBM), \sta, uses information theory to quantify symmetry loss in a finite atomic cluster under a transformation $T_\alpha$. 
The cluster is represented by a normalized electron-weighted density function $\mu$. 
The measure is defined as the Jensen-Shannon (JS) divergence \cite{linDivergenceMeasuresBased1991} between the transformed density $(T_\alpha)_\#\mu$ and the original $\mu$, providing a quantitative assessment of the structure’s deviation from the symmetry element $T_\alpha$. 
The measure can be computed for any operation in any atomic cluster. 
For an undistorted crystal structure, it returns zero for each crystallographic symmetry operation by design. 
However, with finite size clusters whose shape breaks symmetry, unit cell distortions, or atomic displacements, it yields a positive value that indicates the degree of SB with respect to each symmetry operation.

In this paper, we derive this measure and conduct numerical tests to evaluate its behavior for different cases, comparing it with another related SBM based on the Kullback-Leibler (KL). We demonstrate that these measures are useful for investigating factors affecting SB, such as boundary shape, perturbation size, and distorted atoms. They also help identify which symmetries are most violated or preserved when a cluster experience a specific distortion.

\section{KL and JS divergences in symmetry breaking}

Transform Information (TI) has been employed as a quantitative measure of SB, with successful applications in fields such as biological systems~\cite{gandhiIdentificationApproximateSymmetries2021}, and it has been demonstrated to be a general form of many classical information measures~\cite{vstovskyTransformInformationSymmetry1997}. The fundamental idea behind this approach is to compare an object of interest with a transformed version of itself. 
The TI associated with the transformation $T_\alpha$ is defined as
\begin{equation}
\sta = \int_D \mu(\zeta) \ln\left(\frac{\mu(\zeta)}{(T_\alpha)_\#\mu(\zeta)}\right) \, d\zeta,
\end{equation}
where $\mu(\zeta)$ is an intensity function of interest over the domain $D$. The transformation $T$ is parameterized by a continuous variable $\alpha$, such as the angle of rotation around a fixed axis. When $\mu$ is a probability measure, TI becomes a special case of the Kullback-Leibler (KL) divergence~\cite{kullbackInformationSufficiency1951}, also known as relative entropy. The KL divergence, denoted by $D_{\text{KL}}(P\parallel Q)$, quantifies the information loss if one probability distribution $P$ is approximated by another, $Q$. For the case of a continuous random variable, it is defined as
\begin{equation}
D_{\text{KL}}(P\parallel Q)=\int _{\mathcal X} p(x)\log \frac{p(x)}{q(x)}\,dx,
\end{equation}
where $p$ and $q$ denote the probability densities of $P$ and $Q$, respectively, defined on measurable space $\mathcal X$. While the KL divergence possesses the capability to quantify the distance between two probability distributions, it is not mathematically recognized as a valid metric. Specifically, it fails to satisfy the triangle inequality and lacks symmetry. Additionally, it is not always well-defined and could become unbounded. An alternative is the Jensen-Shannon (JS) divergence, given by:
\begin{align}
\begin{split}
D_{\text{JS}}(P, Q) = & \frac 12 D_{\text{KL}}(P\parallel M) + \frac 12 D_{\text{KL}}(Q\parallel M) \\
= & \frac 12\int _{\mathcal X} p(x)\log \frac{p(x)}{m(x)}\,dx + \frac 12 \int _{\mathcal X} q(x)\log \frac{q(x)}{m(x)}\,dx,
\end{split}
\end{align}
where $M$ represents a mixture distribution of $P$ and $Q$, defined as $M=\frac 12 P+\frac 12 Q$. Consequently, the JS divergence is often referred to as the ``total divergence to the average." When the logarithm's base is 2, the JS divergence remains bounded by 1~\cite{linDivergenceMeasuresBased1991}.

Here we study measures of SB based on both the KL and JS divergences and apply them to model systems to understand their behavior.

\section{A symmetry breaking measure for finite clusters}

\subsection{Finite cluster representation}

In our proposed model, we represent a finite cluster of atoms as a normalized electron-weighted atomic density function $\mu(\bm{x}): \bm{x}\in\mathbb R^3\to\mathbb R$. The electron density at each atom is assumed to be located at the position of the atomic nucleus.
However, due to atomic motions, the probability distribution of the atomic density can be approximated by a three-dimensional Gaussian distribution, commonly known as the Debye-Waller approximation \cite{warrenXrayDiffraction1990}, weighted by the number of electrons held by the atom.
Specifically, for any atom $k$ where the average location of the nucleus is at $\bm{x}_k$, the electron-weighted atomic density can be written as
\begin{equation}
\tilde{\mu_k} = e_k\cdot o_k\cdot\mathcal{N}(\bm{x}_k, \bm{U}_k),
\end{equation}
where, $e_k$ denotes the number of electrons held by atom $k$, $o_k$ is its occupancy factor, and $\bm{U}_k\in\mathbb{R}^{3\times 3}$ is the anisotropic atomic displacement tensor (ADT). For simplicity, here we treat this distribution as being isotropic, that is, $\bm{U}_k=U_k\cdot \bm{I}_3$ is a diagonal matrix, and all the diagonal entries, $U_k$, are equal to the \uiso of atom $k$. Here $U_{k} =1/3\times(\bm U_{k,11}+\bm U_{k,22}+\bm U_{k,33})$ can be interpreted as a mean-square displacement averaged over all the three directions~\cite{truebloodAtomicDispacementParameter1996}.


With this representation, the electron-weighted atomic density function of a finite cluster of particles can be expressed as the superposition of Gaussian distributions,
\begin{equation}\label{eq;tilde_mu}
\tilde\mu=\sum_{k=1}^N \tilde{\mu_k} = \sum_{k=1}^N \tilde{\phi_k}\,\mathcal{N}(\bm{x}_k, U_k\cdot  \bm{I}_3),
\end{equation}
where $\tilde{\phi_k}=e_k\cdot o_k$. To transform the electron-weighted atomic density function into a probability density function, we normalize $\tilde\mu$ such that it possesses an $L_1$ norm of unity. This results in the normalized electron-weighted atomic density function,
\begin{equation}\label{eq;mu}
\mu=\frac {\tilde\mu}{\|\tilde\mu\|_{L_1}}=\frac {\tilde\mu}{\sum_{k=1}^N \tilde{\phi_k}} = \sum_{k=1}^N \phi_k\,\mathcal{N}(\bm{x}_k, U_k\cdot  \bm{I}_3),
\end{equation}
where $\phi_k=\tilde{\phi_k}/\sum_{i=1}^N \tilde{\phi_i}$. For the case of isotropic ADTs, a transformation $T_\alpha$ to $\mu$ is equivalently defined by transforming the Gaussian mean of all the particles,
\begin{align}
\begin{split}
(T_\alpha)_\#\mu = & (T_\alpha)_\#\left(\sum_{k=1}^N \phi_k\,\mathcal{N}(\bm{x}_k, U_k\cdot  \bm{I}_3)\right)\\
= & \sum_{k=1}^N \phi_k\,\mathcal{N}(T_\alpha(\bm{x}_k), U_k\cdot  \bm{I}_3). \label{eq;Tamu}
\end{split}
\end{align}
\subsection{Symmetry breaking measure}

Using our finite cluster representation, given a general transformation $T_\alpha$, we first define the SBM based on the KL divergence (KL-SBM), $\sta^{KL}[\mu]$, as
\begin{align}\label{eq;measure_KL}
\begin{split}
\sta^{KL}[\mu]= & D_{\text{KL}}\left(\mu \| (T_\alpha)_\#\mu\right)\\
= & \int _{\mathbb R^3} \mu(\bm{x})\log \left({\frac {\mu(\bm{x})}{(T_\alpha)_\#\mu(\bm{x})}}\right)\,d\bm{x}.
\end{split}
\end{align}
This measure quantifies the similarity of the structure with itself after the transformation of interest $T_\alpha$.
We will explore the performance of this measure in simple cases below.

Although $\sta[\mu]$ is bounded for Gaussian mixtures~\cite{nielsenGuaranteedBoundsInformationTheoretic2016}, the bounding limit is contingent upon specific attributes of the finite cluster, including atom species and the number of atoms incorporated.
As a consequence, generalization to compare the relative SB of dissimilar clusters is not possible.

We also define a SBM based on the JS divergence which we might expect to give a more transferrable measure bound.
We define the the JS-SBM, $\sta^{JS}[\mu]$, as
\begin{equation}\label{eq;measure_JS}
\sta^{JS}[\mu]= \frac 12\int _{\mathcal X} \mu(\bm{x})\log \frac{\mu(\bm{x})}{m(\bm{x})}\,dx + \frac 12 \int _{\mathcal X} (T_\alpha)_\#\mu(\bm{x})\log \frac{(T_\alpha)_\#\mu(\bm{x})}{m(\bm{x})}\,d\bm{x},
\end{equation}
where $m$ is an equal mixture of $\mu$ and $(T_\alpha)_\#\mu$. 
Specifically,
\begin{equation}
m = \sum_{k=1}^N \frac {\phi_k}{2}\,\left[\mathcal{N}(\bm{x}_k, U_k\cdot \bm{I}_3) + \mathcal{N}(T_\alpha(\bm{x}_k), U_k\cdot \bm{I}_3) \right].
\end{equation} 
Although the pointwise log‑ratio can be negative, the integral is equal to a JS-divergence between $\mu$ and $(T_\alpha)_\#\mu$ and is therefore non‑negative by Gibbs' inequality \cite{linDivergenceMeasuresBased1991}. 
$\sta^{KL}[\mu]$ and $\sta^{JS}[\mu]$ both satisfy the three divergence properties:
\begin{itemize}
\item Self similarity: $\sta[\mu]=0$ if $T_\alpha$ is the identity.
\item Self identification: $\sta[\mu]=0$ only if $(T_\alpha)_\#\mu=\mu$.
\item Positivity: $\sta[\mu]\geq 0$ for all $\mu$, given any transformation $T_\alpha$. 
\end{itemize}
The first two properties can be succinctly stated as $\sta[\mu]=0$ if and only if the $T_\alpha$ transformation is a symmetry preserving operation, $\mathscr{O}$, of the structure. In other words,
\begin{equation}
\sbm_\mathscr{O}[\mu] = 0.
\end{equation}
In all other cases, $\sta[\mu]$ is positive.

For the case of the JS-SBM, when using a logarithm with a base of 2 in Eq.~\ref{eq;measure_JS}, it is bounded by one.
If the natural logarithm is employed, the upper bound becomes $\sqrt 2$.
$\sta^{JS}[\mu]$ reaches its upper bound if and only if $\mu$ and $T_\alpha$ are disjoint, indicating that the operation $T_\alpha$ is broken completely. For a Gaussian mixture $\mu$, which is consistently non-zero, this upper bound will not be realized. However, if $\bm{x}_k$ and all $T_\alpha(\bm{x}_k)$ are sufficiently distant in terms of their \uiso's, $\sta^{JS}[\mu]$ can approach close to its bound.

A small but non-zero $\sta[\mu]$ indicates the transformation $T_\alpha$ is only weakly breaking the symmetry.
In this context, $\sta[\mu]$ is a continuous measure of structural SB.
If a structure undergoes a small distortion, for example, due to the displacement of one or several atoms off crystallographic special positions by a small magnitude, the SBM can be calculated for each symmetry operation of the undistorted structure to identify the symmetry operations that have been more severely disrupted, those that remain intact, and those that have been marginally affected.

For a high-symmetry heterostructure, such as an infinite crystal with a Face-Centered Cubic (FCC) lattice of nickel, we refer to a set of symmetry operations denoted as $\mathscr T=\{T_{i}\}$, which contains transformations that, when applied, yield an identical object. We then introduce some changes to this heterostructure. This results in the object losing some of the symmetry operations from the set $\mathscr T$ (and in principle, possibly gaining some new ones) and having a new set of symmetry operations, $\mathscr T'=\{T'_{i}\}$. Strictly speaking, SB of this structure can be defined as the condition where the SBM $\sbm_{T_i}$ of some symmetry operation $T_i \in \mathscr T$ result in $\sbm_{T_i}>0$.

\subsubsection{Symmetry breaking measure of symmetry operators}

The JS-SBM and KL-SBM are formulated as continuous functions of transformations $T_\alpha$, such as $R_\alpha$, which represents a counterclockwise rotation of angle $\alpha$. The SBM is uniquely defined for the transformation and $T_\alpha$ does not need to be a symmetry operation, which is defined as an action leaving an object unchanged.

Determining the SBM of \textit{symmetry operators} is not trivial, because the \textit{symmetry operators} can generate more than one symmetry operation. For example, such as the proper axes of rotation $C_n$ can lead to $n-1$ operations, specifically $C_n,C_n^2,\dots,C_n^{n-1}$. $C_n^n$ is considered the ``identity'' operation, denoted $E$. For example,  in structures with n-fold rotational symmetry $C_n$, the SBM for these operations are zero ($\sbm_{C_n}=\sbm_{C_n^2}=\dots=\sbm_{C_n^{n-1}}=0$). However, if a perturbation disrupts this n-fold symmetry, the SBM for $C_n^m$ and $C_n^k$ may differ for some $m,k\in\{1,\dots,n-1\}$. Therefore, we need a measure to quantify how much a particular symmetry operator, such as n-fold rotation, is broken by a symmetry lowering distortion. 

To address this, we introduce \textit{symmetry breaking measure of operators}, $\sbm_{\mathscr{O}}$, as the average SBM of all the $|\{\mathscr{O}^m\}|$ operations derived from $\mathscr{O}$.
\begin{equation}
\sbm_\mathscr{O}=\frac{1}{|\{\mathscr{O}^m\}|}\sum_{\mathscr{O}^m} \sbm_{\mathscr{O}^m}.
\label{eq;operator_sbm}
\end{equation}
For example $\sbm_{C_4}=1/3 \sum_{m=1}^3 \sbm_{C_4^m}$. Here, a symmetry operator $\mathscr{O}$ includes its geometric element (e.g., for $C_n$ the rotation axis is fixed; for a mirror the plane is fixed). Thus, Eqn~\eqref{eq;operator_sbm} averages over the derived operations $\{\mathscr{O}^m\}$ of this single fixed element (e.g., $C_n^m$ about the same axis) and does not average across different axis placements or mirror–plane altitudes. This measure gives a value of $0$ when symmetry is preserved, but ranges between $0$ and $1$ for the JS-SBM if symmetry is partially broken. 

\subsubsection{Symmetry breaking measure of a single-atom system}\label{sec;single_atom}

To gain insights, we will consider a structure with a single-atom in the unit cell with \uiso$=U$ and which becomes displaced by $\bm{d}$.
WLOG, let 
\begin{align}
\mu= & \mathcal{N}(\bm{0}, U\cdot \bm{I}_3),\, \\
(T_{\bm{d}})_\#\mu= & \mathcal{N}(\bm{d}, U\cdot \bm{I}_3).
\end{align}
In this simple case there is an analytic expression for the the KL-SBM,
\begin{equation}
\sbm^{KL}_{T_{\bm{d}}}= D_{\text{KL}}\left(\mu \| (T_{\bm{d}})_\#\mu\right)
= \frac{d^2}{2U},
\end{equation}
where $d=\|\bm{d}\|_{L_2}$ is the displacement distance. 
Details of the derivation are provided in Appendix~A.
This result suggests that for small distortions $\bm{d}$, the KL-SBM increases with the square of the displacement $d^2$, demonstrating a sensitive measure of even minor positional changes.

\begin{figure}[H]
  	\begin{center}
  	\includegraphics[scale=0.52]{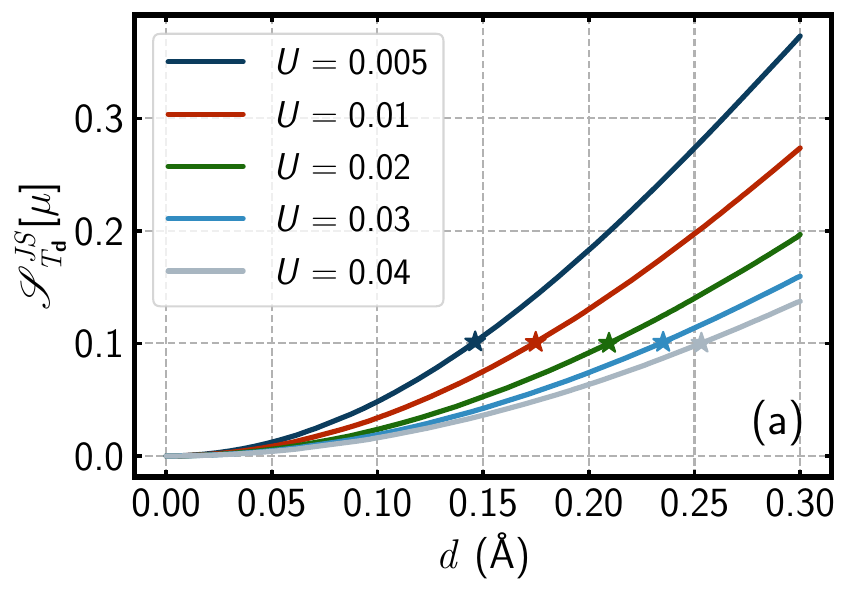}
  	\includegraphics[scale=0.52]{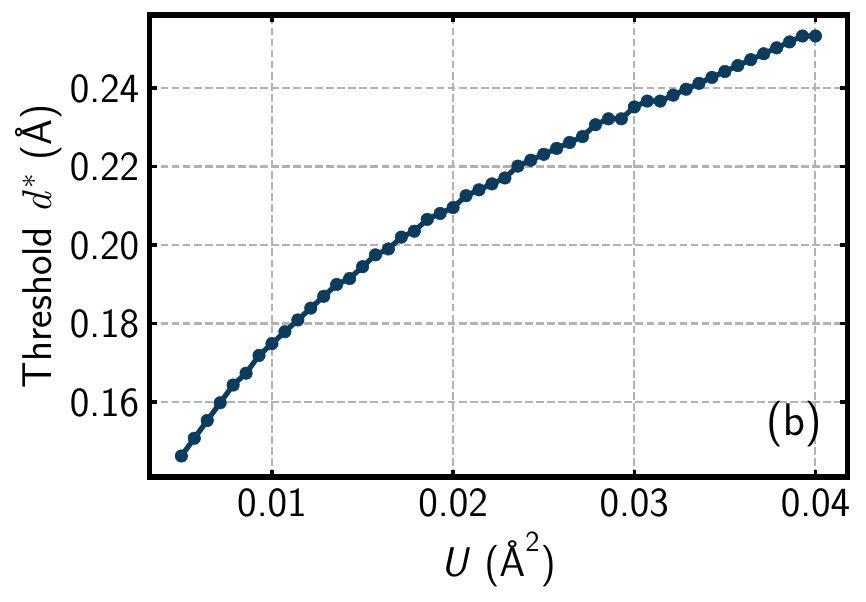}
  	\caption{(a) The JS-SBM of the translational symmetry when an atom is displaced by $\bm{d}$ from itself, plotted for atoms with different \uiso. The star on each curve represents the point at which the JS-SBM reaches $0.1$. (b) The value of $d^*$ where the JS-SBM first reaches the $0.1$ threshold for different \uiso.}
  	\label{fig;two_atom}
  	\end{center}
\end{figure}

The $\sbm^{JS}_{T_{\bm{d}}}[\mu]$ case will be more challenging because it involves a Gaussian mixture $m=\frac{1}{2} \mu + \frac{1}{2} (T_{\bm{d}})_\#\mu$. We compute this numerically and show the result in Figure~\ref{fig;two_atom}(a) for various values of \uiso.  
The JS-SBM increases approximately quadratically with increasing $d$.  
As we might expect, it goes up more slowly when \uiso is larger. 
Panel~(a) focuses on the small-$d$ regime to display $d^{*}$ clearly; for completeness, under logarithm base 2, the JS-SBM curves approach an asymptote $\le 1$ as $d\to\infty$. 
In Figure~\ref{fig;two_atom}(b), we consider a threshold value of $d^*$ at which the JS-SBM reaches $0.1$ and plot this as a function of \uiso$=U$. 
The slope of the plot indicates that achieving a JS-SBM of $0.1$ requires an approximate increase of $0.03$~\AA\ in the displacement magnitude $d$ for every $0.01$~\AA$^2$ increase in $U$.

\section{Numerical Methods}

Here we discuss methods for computing the SBM. Neither the KL or JS divergences have closed-form expressions for Gaussian Mixture Models. In this case, Monte Carlo (MC) simulation emerges as a useful technique for estimating $D_{\text{KL/JS}}(f\|g)$ with arbitrary accuracy~\cite{hersheyApproximatingKullbackLeibler2007}. In this section, our focus is primarily on outlining the method to approximate $\sta^{KL}[\mu]$ and determine the appropriate sample size for the MC simulation, contingent upon a set confidence level and a predetermined tolerance for estimation inaccuracies. The results are relevant also for $\sta^{JS}[\mu]$ which is computed by combining two distinct KL-divergence computations.

The MC simulation expresses the KL divergence as the expectation of $\log(f/g)$, under the probability density function~$f$. In other words,
\begin{align}
\begin{split}
\sta^{KL}[\mu] & = \int _{\mathbb R^3} \mu(\bm{x})\log \frac {\mu(\bm{x})}{(T_\alpha)_\#\mu(\bm{x})}\,d\bm{x}\\
&=  \mathbb E_{\bm{x}\sim\mu} \left[\log \frac {\mu(\bm{x})}{(T_\alpha)_\#\mu(\bm{x})}\right]. \label{eq;MC}
\end{split}
\end{align}
The MC methodology can then be applied to estimate the expectation values using the algorithm,
\begin{algorithm}[H]
\caption{The Monte Carlo estimation of $\sta^{KL}[\mu]$} \label{alg;MC}
\begin{algorithmic}
\State 1. Draw $M$ independent and identically distributed (i.i.d.) samples $\bm{x}_i$ from $\mu$.
\State 2. Compute $\left(\sta^{KL}\right)^{MC}[\mu]=\frac 1 M \sum_{i=1}^M h(\bm{x}_i)$, where $h(\bm{x}_i)=\left[ \log \left({\frac {\mu(\bm{x}_i)}{(T_\alpha)_\#\mu(\bm{x}_i)}}\right)\right]$.
\end{algorithmic}
\end{algorithm}
By the law of large numbers, the MC estimate $\left(\sta^{KL}\right)^{MC}[\mu]$ converges to $\sta^{KL}[\mu]$ as the number of samples $M\to\infty$. The estimation error is of order $O(1/\sqrt M)$. One can construct a confidence interval for the MC estimate as
\begin{align}
CI=\left(\left(\sta^{KL}\right)^{MC}[\mu]-z\frac{\sigma}{\sqrt M},\left(\sta^{KL}\right)^{MC}[\mu]+z\frac{\sigma}{\sqrt M}\right), \label{eq;CI}
\end{align}
where $\sigma$ is the standard deviation of $\{h(x_i)\}_{i=1}^M$, and $z$ is the $z$-score determined by the confidence level. For instance, for a $95\%$ confidence interval, implying a $95\%$ chance of containing the true value of $\sta[\mu]$, $z$ is approximately equal to $1.96$. One can estimate the required sample size $M$ using the following algorithm:
\begin{algorithm}[H]
\caption{Estimation of Monte Carlo sample size $M$} \label{alg;sample_size}
\begin{algorithmic}
\State 1. Choose an arbitrary large $M'$, and draw $M'$ i.i.d. samples $\bm{x}_i$ from~$\mu$.
\State 2. Compute $\sigma^2=\text{Var}(h(\bm{x}_i))$, where $h(\bm{x}_i)=\left[ \log \left({\frac {\mu(\bm{x}_i)}{(T_\alpha)_\#\mu(\bm{x}_i)}}\right)\right]$.
\State 3. The recommended sample size $M$ is $(z\cdot \sigma/\epsilon)^2$, where $\epsilon$ is the error tolerance on each side.
\end{algorithmic}
\end{algorithm}
For example, the minimum sample size that ensures a $95\%$ probability of the true $\sta^{KL}[\mu]$ being within $CI=(\left(\sta^{KL}\right)^{MC}[\mu]\pm\epsilon)$ is $(1.96\cdot \sigma/\epsilon)^2$. Increasing the value of $M'$ in Algorithm \ref{alg;sample_size} improves the accuracy of the estimated standard deviation $\sigma$, thereby leading to a more precise estimation of the sample size $M$.

The JS divergence, which averages two KL-divergences, can be calculated following Algorithm \ref{alg;MC} by applying it separately to each component. Similarly, the confidence interval and required sample size for each estimation can be derived using Algorithm \ref{alg;sample_size}. Notably, due to the interdependence of the two distributions, the confidence interval for their combined sum cannot be obtained by simply merging the individual confidence intervals.

\section{Numerical Results}

This section discusses how the SBM quantitatively analyzes distortions in finite clusters. When focusing on local distortions, we often examine smaller, locally distorted segments within larger structures. Here, we present two finite clusters as examples, each a segment cut from infinite crystal structures, to demonstrate the practical use of SBM for local distortion analysis. The first cluster is taken from an FCC Nickel crystal structure. Its high inherent symmetry makes it ideal for evaluating how SBM detects symmetry violations. We also examine how different structural cutout choices introduce SB.
The second example involves a supercell derived from a distorted perovskite with octahedral tilts. This case illustrates how SBM enables dynamic analysis of distortion processes by tracking SBM as the perovskite transitions from its original, undistorted state to one with octahedral tilts.

\subsection{The violation of symmetry elements of local distorted Nickel}\label{sec;nickel}

\subsubsection{The symmetry breaking from the boundary of the finite cluster} \label{sec;boundary}

Here we explore how the SBM behaves under different situations. We begin with finite clusters of atoms that are cut out from larger bulk crystals. This simulates idealized nanoparticles where there are no local atomic displacements or relaxations except for the finite size of the particle. This is not different from a point-group symmetry analysis of discrete molecules. However, we are interested in this as an illustrative example of quantifying the SBM inherent in the nanosizing.

\paragraph{Test Design}
In our exploration, we create finite chunks of material where the particle shape either preserves or breaks the underlying symmetry, and we delve into each scenario. We investigate a counterclockwise rotation operation $R_\alpha$ along the 4-fold symmetry axis of the face-centered cubic (FCC) nickel structure, considering spherical and cubic cutouts (point-symmetry-preserving), as well as spheroidal and rectangular cutouts (symmetry-lowering):
\begin{itemize}
\item \textbf{Spherical Cutout}: Contains all atoms within a distance of one lattice parameter from the central nickel atom.
\item \textbf{Cubic Solid Cutout}: Has a side length of one lattice parameter, forming a regular unit cell of nickel.
\item \textbf{Spheroidal Cutout}: Centered on a central atom, oriented with the unique long axis in the crystallographic $ab$-plane, extending two unit cells along the in-plane major axis, one unit cell along the orthogonal in-plane minor axis, and one unit cell along the $c$-axis, containing all atoms within this volume.
\item \textbf{Rectangular Solid Cutout}: Encompasses two adjacent unit cells.
\end{itemize}
The rotation axis passes through either the central nickel atom or the cluster’s center of mass (if no central atom exists) and is parallel to the positive $c$-axis of the original crystallographic unit cell.

In this test, the lattice parameter for nickel was chosen to be $3.52$~\AA.
The atomic displacement parameters (ADP) for each nickel atom were set to \uiso$ = 0.013$~\AA$^{2}$.
For each nickel atom, the electron count is 28, which aligns with its atomic number, $Z$. Additionally, the occupancy, $o$, is set to one for each site. For the evaluation of SBM, both $\sbm^{KL}_{R_\alpha}[\mu]$ and $\sbm^{JS}_{R_\alpha}[\mu]$ were estimated using Monte Carlo random sampling. The sample sizes for these estimations were determined based on a 95\% confidence interval, with a bilateral error tolerance set at $0.025$ for the KL-SBM and $0.0025$ for the JS-SBM.

\begin{figure}[h]
  	\begin{center}
  	\includegraphics[scale=0.5]{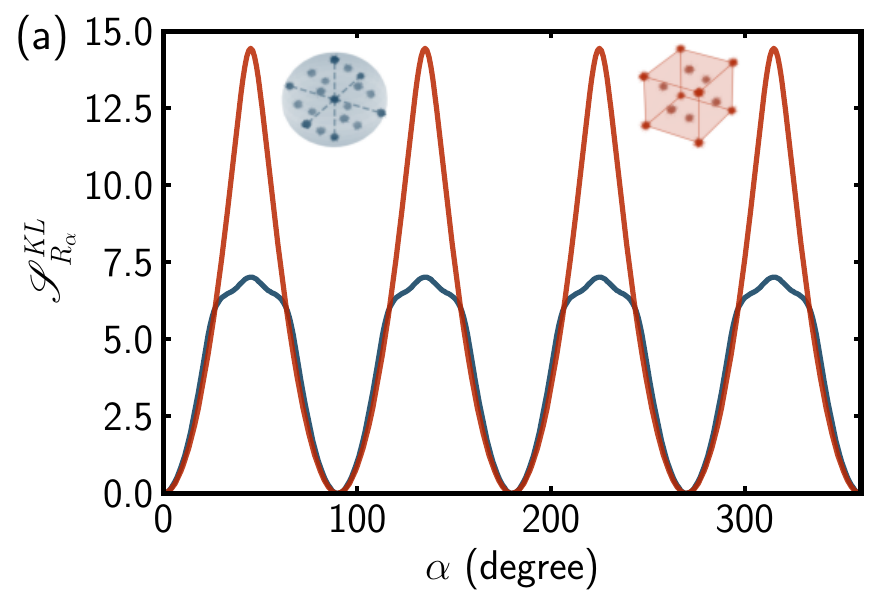}
  	\includegraphics[scale=0.5]{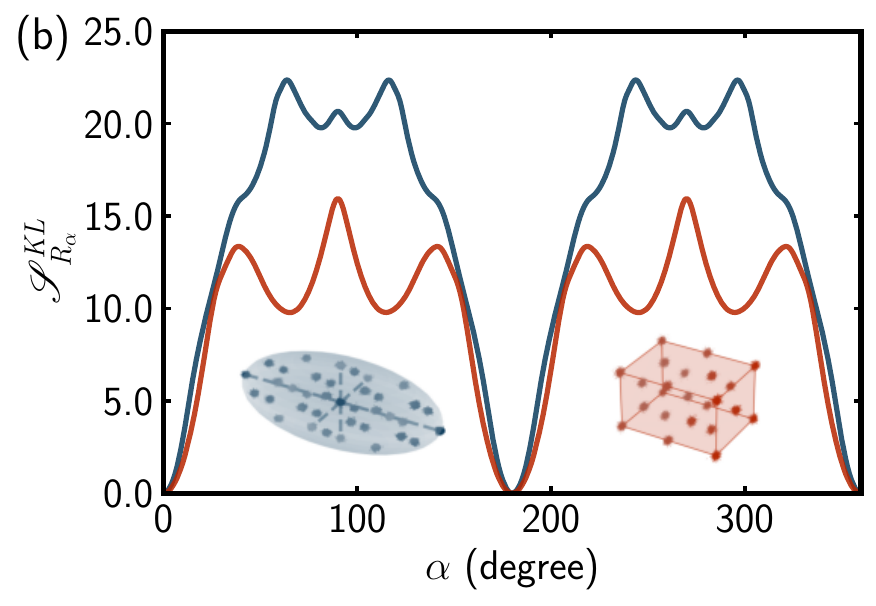}\\
  	\includegraphics[scale=0.5]{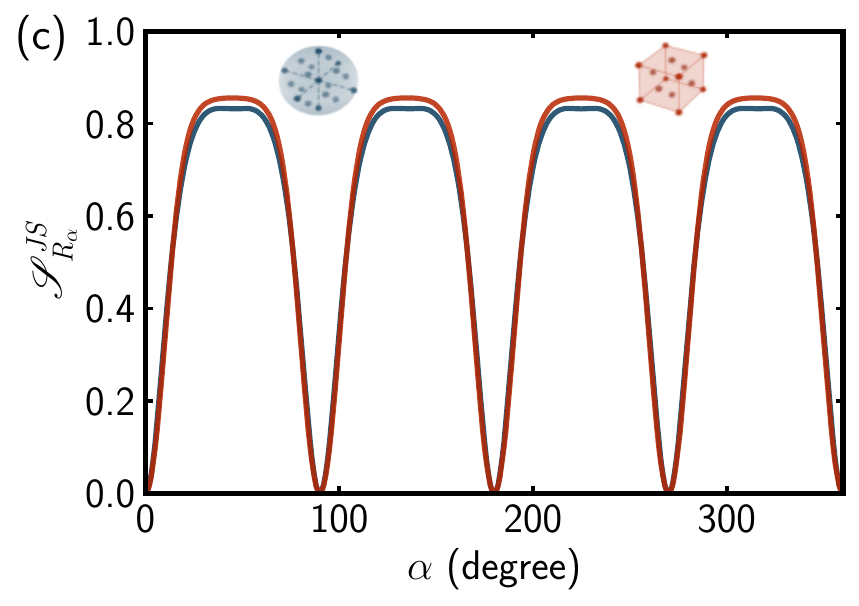}
  	\includegraphics[scale=0.5]{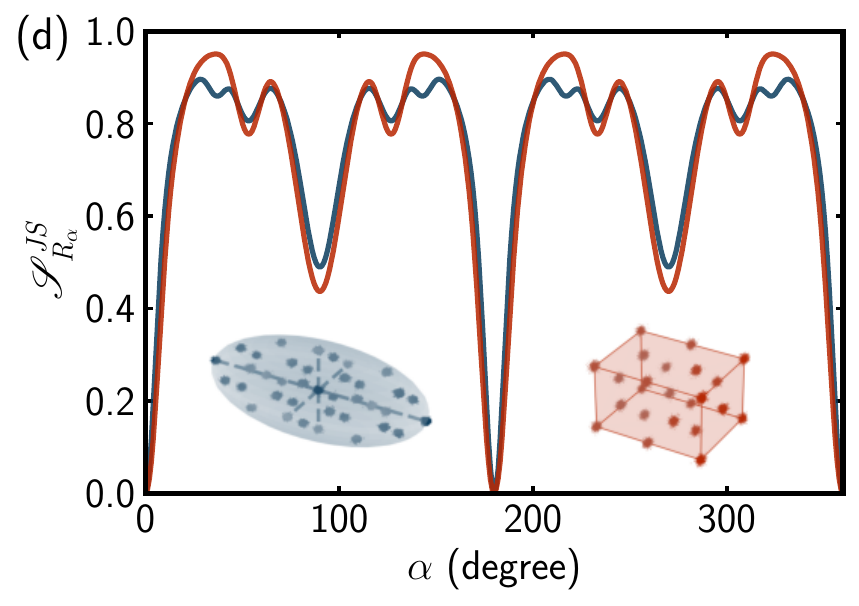}
  	\caption{(a and b) The KL-SBM $\sbm^{KL}_{R_\alpha}[\mu]$ and (c and d) the JL-SBM $\sbm^{JS}_{R_\alpha}[\mu]$, extracted from finite clusters of a nickel crystal structure. In subplots (a and c), spherical (blue) and cubic (red) shapes are used for cutouts, while spheroidal (blue) and rectangular (red) shapes are demonstrated in subplots (b and d). These curves depict the variation of $\sbm_{R_\alpha}[\mu]$ as the respective cluster is rotated by angle $\alpha$ (measured in degrees). This rotation is about an axis aligned with the crystallographic $c$-axis and intersects the cluster's center of mass.}
  	\label{fig;sbm_rotation}
  	\end{center}
\end{figure}
\paragraph{SBM for symmetry-preserving cutouts}
We first consider the point-symmetry-preserving cutouts in Fig.~\ref{fig;sbm_rotation} (a and c). By definition, both KL-SBM and JS-SBM $\sbm_{R_\alpha}[\mu]=0$ when $\alpha=0$ as the object is compared with itself. As we rotate the angle away from zero, $\sbm_{R_\alpha}[\mu]$ increases. Because the object has a four-fold rotational symmetry, $C_4$, for rotations about this axis, we expect both SBM $\sbm_{R_\alpha}[\mu]=\sbm_{C_4^{m}}[\mu]=0$ when $\alpha=m\pi/2$~rad for $m\in\{1,2,3\}$, which results in $\sbm_{R_\alpha}$ being a function with 4-fold periodicity in $\alpha$, as is seen. The fact that these cutouts do not break the point-symmetry is evident in the fact that $\sbm_{R_\alpha}=0$ for every $90$ degrees of rotation, preserving the 4-fold rotational symmetry present in the underlying structure.

The two symmetry-preserving cutouts result in similar $\sbm^{KL}_{R_\alpha}[\mu]$ trajectories for small values of $\alpha$ away from the high symmetry points. The curves initially increase at about the same rate. Both curves go through a maximum at 45$^\circ$, though the cubic cutout has a much higher maximum at this point. The sphere has a higher overall symmetry and this seems to result in a lower maximum of KL-SBM.

The behavior of $\sbm^{JS}_{R_\alpha}[\mu]$ is also similar for each of these clusters at small deviations of $\alpha$ from high symmetry points.  However, the behavior of JS-SBM is notably different in the proximity of their local maxima where the two curves follow each other closely. Both trajectories rapidly approach their local maxima, approximating a value of $0.85$, and subsequently remain relatively constant around the 45$^\circ$ rotation.

\paragraph{SBM for symmetry-lowering cutouts}

In the cases of spheroid and rectangular solid cutouts, the internal structure does not deviate from the ideal FCC nickel structure, but the shape of the cluster takes the object from 4-fold to 2-fold symmetric about this rotation axis. We therefore expect the SBM to be a 2-fold periodic function with zeros at 0 and 180 degrees of rotation. This is indeed what is seen in Fig.~\ref{fig;sbm_rotation} (b and d).

Analogous to the symmetry-preserving cases, the profiles of $\sbm^{KL}_{R_\alpha}[\mu]$ for these cutouts display pronounced discrepancies. Interestingly, in both cases, the 90$^\circ$ rotation which was a minimum and went to zero in the symmetry-preserving cutouts, is now a local maximum in the KL-SBM, and for the rectangular cutout is a global maximum.
This might be expected from the point of view that the elongated structures are perpendicular at this point.  However, on the other hand, the underlying structures will come into coincidence again at these points for the atoms that do overlap due to the internal four-fold symmetry of the structure, so it is not so obvious how we might expect a SBM to behave.

On the other hand, the behavior of $\sbm^{JS}_{R_\alpha}[\mu]$ for the spheroidal and rectangular solid cutouts exhibits much more congruence. Both cutouts yield trajectories that are closely aligned with each other.  In this case, the 90$^\circ$ rotation results in a local minimum in this measure.  As expected, it does not go to zero, but for both cutouts, it is a local minimum with a value of around 0.5. The nonzero local minimum at 90$^\circ$ and 270$^\circ$ for JS-SBM suggest that while the cutouts disrupt $C_4$ symmetry, the interior Nickel structure retains this symmetry to some extent.

\paragraph{KL-SBM $\sbm^{KL}_{R_\alpha}[\mu]$ v.s. JS-SBM $\sbm^{JS}_{R_\alpha}[\mu]$}
This observation suggests that JS-SBM is less sensitive to the boundary characteristics and particle quantity within the finite cluster. Additionally, the fact that JS-SBM values range between $0$ and $1$ facilitates direct comparison of SB across different clusters, making JS-SBM a more standardized metric. Consequently, the JS-SBM's consistent and straightforward behavior makes it more suitable for comparative analysis of SB between distinct clusters compared to KL-SBM.
Given these insights, the subsequent focus will exclusively be on the JS-SBM.

\paragraph{SBM of \textit{symmetry operators}}
By an inspection of the four-fold rotational symmetry, $C_4$ was not broken at all by the cubic and spherical cutouts but it was broken by the spheroidal and rectangular cutouts.
For the sphere and cube, we find that
\begin{equation}
\sbm_{C_4}=\frac{1}{3}\sum_{m=1}^{3}\sbm_{C_4^{m}}
=\frac{1}{3}\sum_{m=1}^{3}\sbm_{R_{m\pi/2}}=0,
\end{equation}
so our operator SBM correctly returns zero as expected.
We find that for the spheroid, $\sbm^{JS}_{C_4}\approx 0.33$ and for the rectangular solid cutout, $\sbm^{JS}_{C_4}\approx 0.29$. Thus, by the measure $\sbm^{JS}_{C_4}$, we may argue that for this example, the spheroid cutout breaks the symmetry approximately the same as, but slightly more strongly than, the rectangular solid cutout.

\subsubsection{The symmetry breaking from local perturbations}\label{sec;nickle_test_2}

We now investigate the different ways that displaced atoms contribute to SB. As an illustrative example, we will use the same cubic cutout from the Nickel structure. Unless otherwise noted, we use the same isotropic ADP as above, \uiso$=0.013$~\AA$^{2}$, for all nickel atoms.

\paragraph{Deviation from four-fold rotational symmetry $C_4$}

\begin{figure}[!ht]
  \centering
  \begin{minipage}[t]{0.48\linewidth}
    \centering
    \includegraphics[width=\linewidth]{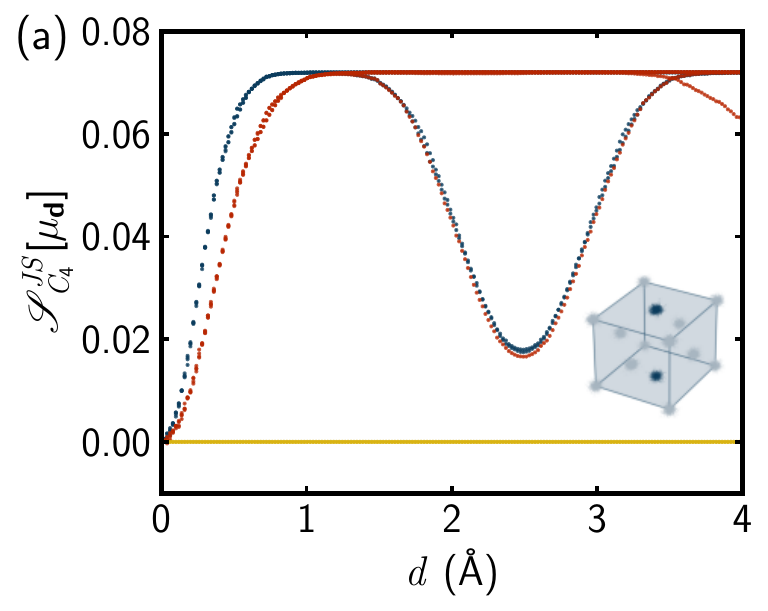}
  \end{minipage}\hfill
  \begin{minipage}[t]{0.48\linewidth}
    \centering
    \includegraphics[width=\linewidth]{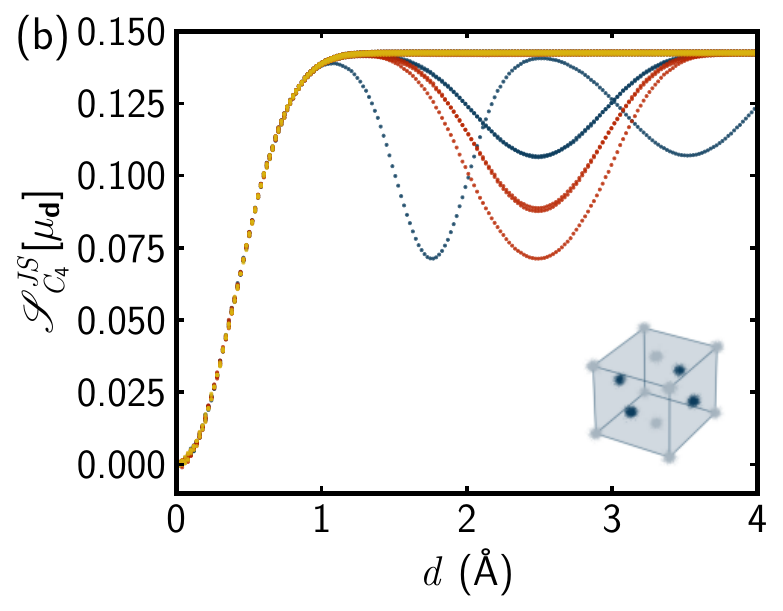}
  \end{minipage}
  \caption{JS–SBM $\sbm^{JS}_{C_4}[\mu_{\vec d}]$ vs. displacement magnitude $d=|\vec d|$. Panel a (left): displacing a top/bottom face atom on the $C_4$ axis. Colors encode displacement direction: blue (along $a$-$b$ plane), red ($45^{\circ}$), yellow (along $c$). Panel b (right): displacing a side-face atom off-axis (same color coding). The $C_4$ axis is fixed through the cluster center.}
  \label{fig;rotation_distortion}
\end{figure}

We first test the four-fold rotational symmetry element $C_4$ that goes through the center of the cluster and is parallel to $c$, as previously discussed.
Perturbations are introduced within the finite cluster by selecting a single atom and displacing it with a vector $\vec{d}$.
We then compute the JS-SBM, $\sbm^{JS}_{C_4}[\mu_{\vec{d}}]$. The results are summarized in Fig.~\ref{fig;rotation_distortion} plotted as a function of the magnitude of $\vec{d}$, $d = |\vec{d}|$. To guide the interpretation of Fig.~\ref{fig;rotation_distortion}, we consider three displacement-vector scenarios (in both panels): 
\begin{itemize}
    \item within the $a$-$b$ plane (shown in blue) with two representative directions $(1,0,0)$ and $(\sqrt 2/2,\sqrt 2/2,0)$,
    \item at a $45^{\circ}$ angle to the $a$-$b$ plane (shown in red) with two representative directions $(\sqrt 2/2,0,\sqrt 2/2)$ and $(1/2,1/2,\sqrt 2/2)$,
    \item parallel to $c$ (shown in yellow).
\end{itemize}
Panel a (left) shows displacements of a top/bottom face atom on the $C_4$ axis, whereas Panel b (right) shows displacements of a side-face atom off the axis.

We first note that the magnitudes of $\sbm_{C_4}[\mu]$ are considerably smaller than those from the rigid rotation of the magnitudes shown in Fig.~\ref{fig;sbm_rotation}.
This observation is reasonable since only one particle out of the ensemble of $14$ undergoes displacement, while the others remain unchanged.

Displacing an atom lying on the rotation axis in a direction along the axis (top or bottom centering atom displaced along $c$) does not break symmetry and $\sbm^{JS}_{C_4}[\mu_{\vec{d}}]=0$ for all $d$ (yellow curve in Panel a of Fig.~\ref{fig;rotation_distortion}) as expected.

Displacing these same atoms with a component perpendicular to the rotation axis does break symmetry. 
$\sbm^{JS}_{C_4}[\mu_{\vec{d}}]$ increases smoothly as shown by the blue and red curves in Panel a of Fig.~\ref{fig;rotation_distortion}.  
It saturates to a value of $0.072$ for all displacement directions.
This value corresponds to the atomic density of a single atom becoming effectively unmatched under the $C_4$ operation (i.e., $\approx 1/14$, since we have 14 atoms in the cluster). This plateau is reached when the atom moves far enough from its original position that the overlap under $C_4$ rotation is negligible. 

For displacements of atoms which are on the non-axial face centers, the SBM plateau is at $0.143$ ($\approx 2/14$).
In this case the displaced atom becomes effectively unmatched with respect to both its own rotated copy and a neighboring site, rather than just itself (see Panel b).

All the curves smoothly increase from $0$ with increasing $d$ and eventually reach a plateau in both panel a and b. 
To gain insight into the behavior of the SBM, consider the blue curve in Panel a: the atom on the rotation axis is displaced in the $a$-$b$ plane by a distance $d$. 
After applying the $C_4$ symmetry operation to the displaced atom, the separation between the original and the rotated atom is $\sqrt{2}d$.
The SBM approaches its plateau once the displaced but unrotated atomic density has a negligible overlap with its $C_4$-rotated counterpart. For Nickel’s isotropic ADP $U_{\mathrm{iso}}=0.013$~\AA$^{2}$, the single-atom translation case reaches near-saturation (JS–SBM $\gtrsim 0.95$) at a displacement of about $0.93$~\AA\ in our simulation. In the present $C_4$ geometry the separation between those two positions is $\sqrt{2}\,d$, so the corresponding near-saturation in Panel~a (blue curve) occurs at $d \approx 0.93/\sqrt{2} = 0.66$~\AA.
This numerical value $0.93$~\AA\ is specific to $U_{\mathrm{iso}}=0.013$~\AA$^{2}$. Larger $U_{\mathrm{iso}}$ requires a larger displacement to reach the same near-saturation level because overlap decays more slowly.

Some of the atoms remain effectively non-overlapping under $C_4$ with increasing $d$ and $\sbm^{JS}_{C_4}[\mu_{\vec{d}}]$ stays on the plateau; for example, the red and blue curves in Panel a for the $(1,0,0)$ displacement direction.
However, in other cases, $\sbm^{JS}_{C_4}[\mu_{\vec{d}}]$ becomes reduced with increasing $d$, an apparently paradoxical result that a larger distortion results in a lower SB.
The reason for this is that the displacement of the atom is so large that it starts to regain appreciable overlap with another atom in the structure. 
For example, the displacement direction of $(1,1,0)$ of the atom at the center of the top face when it approaches the corner of the cube.
Logically and mathematically this makes sense, though this does not correspond to a real situation that would be encountered in practice.
However, plotting $\sbm^{JS}_{C_4}[\mu_{\vec{d}}]$ over such an unphysically large range helps us to build intuition about the function.

\paragraph{Deviation from the reflection operator $\sigma_h$}


\begin{figure}[!ht]
  	\begin{center}
  	\includegraphics[scale=0.7]{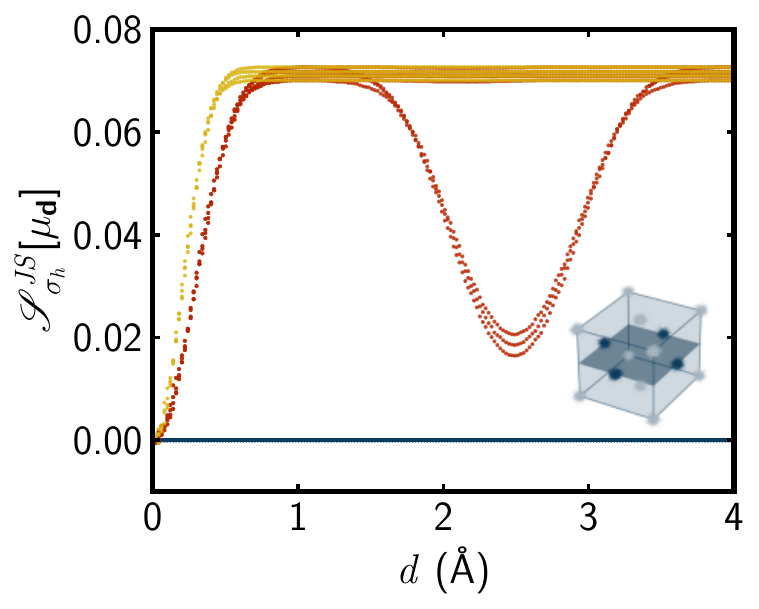}
  	\end{center}
\caption{The SBM $\sbm^{JS}_{\sigma_h}[\mu_{\vec{d}}]$ of a cubic cutout from the Nickel structure when one of its in-plane atoms (darker blue atoms) is shifted parallel to the plane (blue), deviated from the plane by 45 degrees (red), or perpendicular to the plane (yellow). The SBM is calculated for the reflection operation $\sigma_h$, whose mirror plane passes through the center of the finite cluster, with its normal vector along the positive $z$-axis. The figure is plotted as a function of the length of the atom displacement $d$.} \label{fig;reflection_distortion}
\end{figure}

We now consider SBM changes for the reflection operation $\sigma_h$, where the mirror plane passes through the center of mass of the finite cluster and its normal vector points towards the positive $c$-axis.

For an atom lying in the reflection plane atom, when the displacement is parallel to the mirror plane, $\sbm_{\sigma_h}[\mu_{\vec{d}}] = 0$ regardless of the displacement magnitude $d$, which is consistent with the blue curve in Fig.~\ref{fig;reflection_distortion}.

In contrast, when displacing an in-plane atom orthogonally to the mirror plane (as depicted by the yellow curve), $\sbm^{JS}_{\sigma_h}[\mu_{\vec{d}}]$ increases and rapidly converges, reaching a value of approximately $0.072$ when $d=0.60$~\AA. At this displacement, the atomic density and its reflection are effectively disjoint. 

When considering a displacement vector angled at $45$ degrees from the plane (as illustrated by the red curve), the $\sbm^{JS}_{\sigma_h}[\mu_{\vec{d}}]$ primarily depends on the component of the displacement vector normal to the plane, $d_{\perp}=d\cos\theta$, where $\theta$ is the angle between the displacement vector and the normal. Consequently, the curve rises more slowly compared to the orthogonal case.

\begin{figure}[!ht]
  	\begin{center}
  	\includegraphics[scale=0.7]{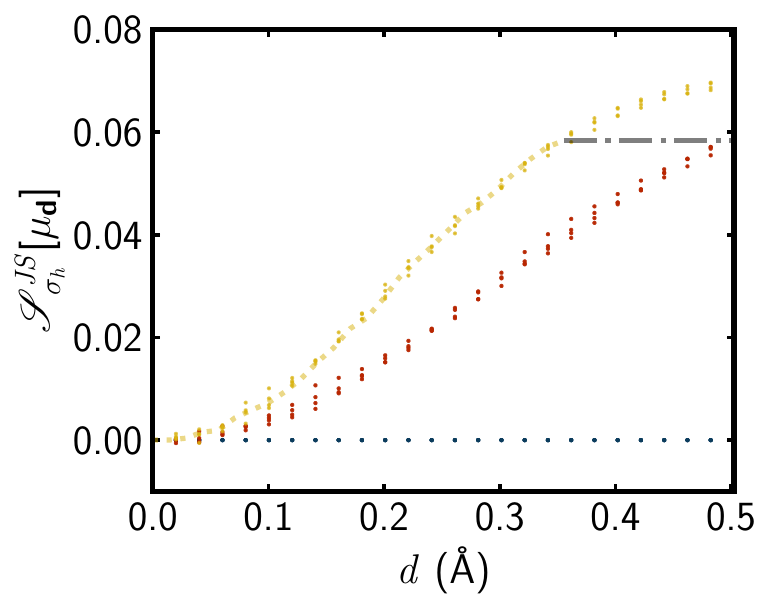}
  	\end{center}
\caption{The SBM $\sbm^{JS}_{\sigma_h}[\mu_{\vec{d}}]$ of a cubic cutout from the Nickel structure calculated for the reflection operation $\sigma_h$ (mirror plane normal to positive $c$-axis). The measure is plotted against displacement magnitude $d$ for shifts parallel to the plane (blue), perpendicular to the plane (yellow), and at 45 degrees (red). The yellow curve (left) represents the perpendicular displacement. It coincides with the trajectory obtained by scaling the $d$-axis of the 45-degree red curve (right) by a factor of $\cos(45^\circ) = 1/\sqrt{2}$, confirming that the scaled angled displacement is equivalent to the perpendicular displacement.} \label{fig:reflection_sqrt2}
\end{figure}

The behavior of $\sbm^{JS}_{\sigma_h}[\mu_{\vec{d}}]$ for small displacements is shown in \fig{reflection_sqrt2}.
In this figure we also show the curves obtained by multiplying $d$ by $\cos\theta = 1/\sqrt{2}$.  
The scaled curves lie on top of each other as expected, showing that it is just the $c$ component of the displacement that contributes to the SB.

As before, some atoms exhibit the anomalous reduction in SBM at large displacements. For instance, when an in-plane atom is displaced towards and overlaps one of its closest corner atoms, depending on the displacement vector, a local minimum is observed at $d=2.49$~\AA, characterized by $\sbm^{JS}_{\sigma_h}[\mu_{\vec{d}}]=0.018$.

\begin{figure}[!ht]
  	\begin{center}
  	\includegraphics[scale=0.55]{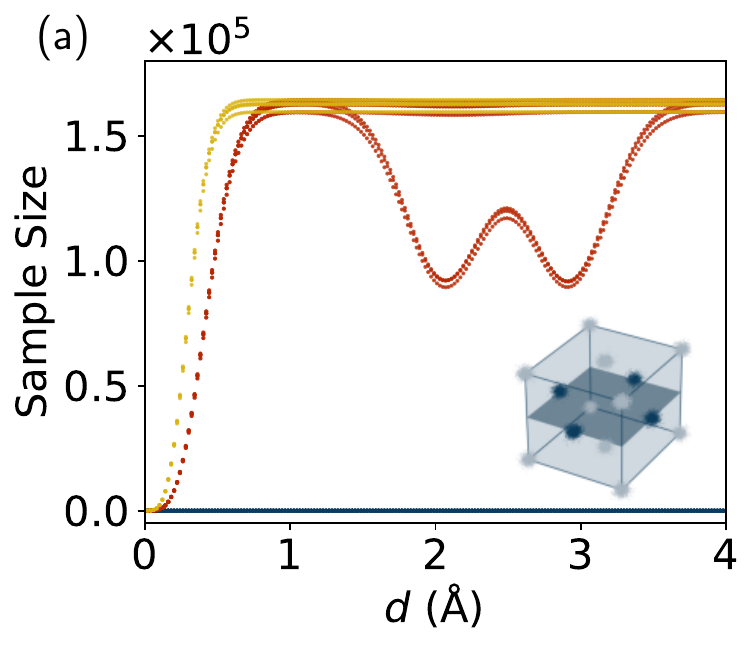}
  	\includegraphics[scale=0.55]{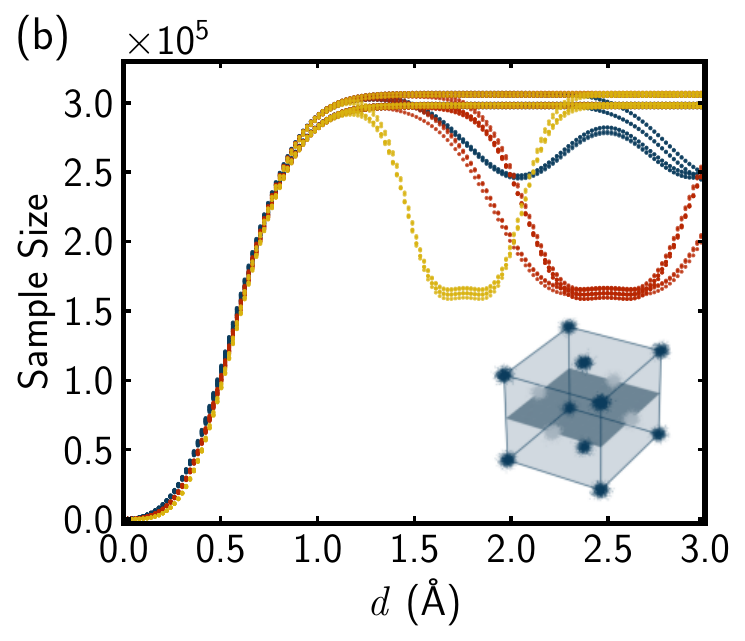}
  	\caption{The recommended sample size for Monte Carlo simulations of $\sbm^{JS}_{\sigma_h}[\mu]$.
  The sample size for (a) finite clusters with displacements of in-plane atoms and (b)  for displacements of out-of-plane atoms. The colors of the curves indicate the sample size for displacements parallel to the plane (blue), 45 degrees to the plane (red), and perpendicular to the plane (yellow). These were the sample sizes needed to give $\sbm^{JS}_{\sigma_h}[\mu]$ with a $95\%$ confidence interval and an error tolerance of $.0025$ on each side.}
  	\label{fig;reflection_samplesize_js}
  	\end{center}
\end{figure}

\paragraph{Monte Carlo sample size analysis}
We make a note here to highlight certain numerical intricacies encountered during our tests. As discussed, we employ a Monte Carlo simulation to estimate the SBM. We found that the sample size required to obtain an estimate of the SBM at a certain level of accuracy increases as $(T_\alpha)_\#\mu$ deviates further from $\mu$. This implies that as atomic displacements magnify, Monte Carlo calculations demand more extensive sample sizes to yield estimates of SBM with consistent precision. In Fig.~\ref{fig;reflection_samplesize_js}, we plot the sample sizes that were used to estimate $\sbm^{JS}_{\sigma_h}[\mu]$. These sizes were determined by considering a $95\%$ confidence interval and a bilateral error tolerance of $0.0025$. The procedure for computing the sample size is delineated in Algorithm~\ref{alg;sample_size}. All the simulation tests addressed in this section employ a sample size determined through this methodology.

\subsection{The violation of symmetry elements of local distorted perovskites from rotations of rigid units} \label{sec;perovskites}

Here we consider a SB by a collective displacement of multiple atoms. For example, this might include the case of a second order structural phase transition due to a soft phonon mode.

\paragraph{Test Design}
For a concrete case, we consider the perovskites, a material class with nominal stoichiometry ABX$_3$. Due to their structural geometry that is well approximated as corner-shared rigid octahedra, the octahedra can collectively tilt in several different patterns, which can be described using a $2 \times 2 \times 2$ (or smaller) supercell of the cubic perovskite unit cell, as described in Glazer's classification~\cite{glazerClassificationTiltedOctahedra1972} of allowed tilt patterns. For simplicity, we consider here a tilt system where an octahedron has no tilt around the $a$ and $b$ axes and only allows for a non-zero in-phase tilt around the $c$ axis, corresponding to the tilt pattern $a^0a^0c^+$ (No. 21 tilt system) in Glazer's classification~\cite{glazerClassificationTiltedOctahedra1972}. However, it serves our purpose as it allows us to explore the effect of collective rotations on the SBM.

The collective rotations are modeled using algebraic expressions that link displacements of atoms so as to preserve the rigid linked octahedral rotations. Note that to maintain corner-connectivity, this distortion couples the tilt angle to the unit cell dimensions; consequently, the lattice parameters change, resulting in a non-cubic supercell and inducing positional shifts for all atoms in the finite cluster relative to the reference frame. We have employed similar algebraic expressions previously for data simulation~\cite{skjaervoRefiningPerovskiteStructures2022}. 

\begin{figure}[!ht]
  \includegraphics[width=\linewidth]{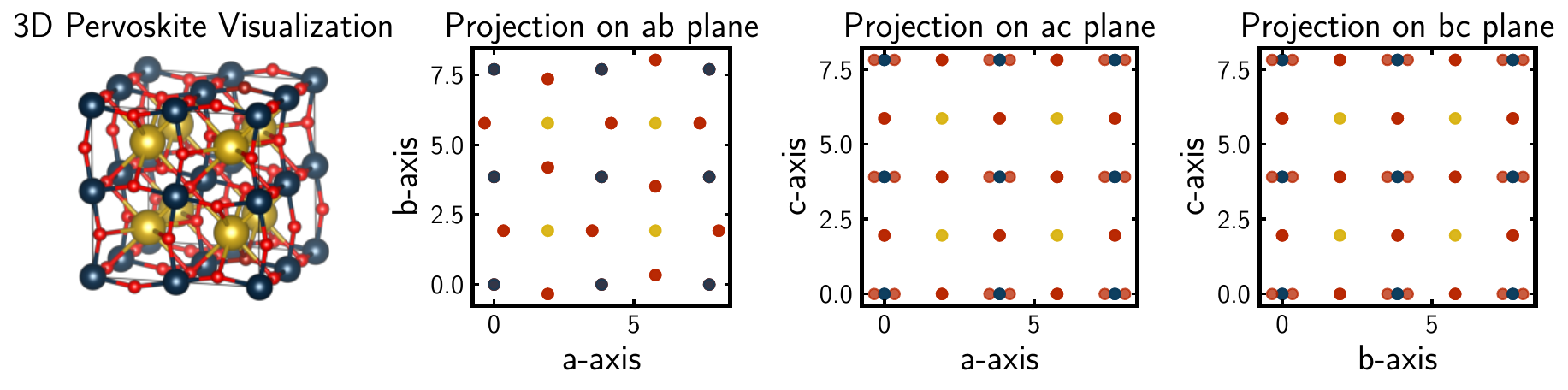}
\caption{Visualization of an in-phase tilt systems projected along the three crystallographic axes in the $2\times 2\times 2$ supercell, with a tilt angle of $\phi =10^{\circ}$. 
The distortion corresponds to the $a^0a^0c^+$ (No.~21 tilt system) in Glazer's classification. 
In the representation, Ca cations are yellow, Ti cations are blue, and O are red. All dots are uniformly scaled to enhance clarity in the spatial distribution of the cations.}
\label{fig;GM21_10}
\end{figure}

Specifically we consider CaTiO$_3$ and use crystallographically reasonable $U_{iso}$ values of $0.0052$~\AA$^2$, $0.0027$~\AA$^2$, and $0.0104$~\AA$^2$ for Ca, Ti and O, respectively. The lattice parameter of the undistorted cubic perovskite is $3.91$~\AA.
Illustrations of this in-phase tilt pattern, as viewed down each tilt axis, are shown in Figure~\ref{fig;GM21_10}.

\paragraph{SBM of the $C_4$, symmetry operator}
As this distortion process occurs, the space group changes from $Pm\bar{3}m$ (No.~221) to $P4/mbm$ (No.~127). Several symmetry operations that the tilted perovskite $\mu_\phi$ breaks and preserves are tested, and the evolution of the SBM $\sbm^{JS}_\mathscr{O}[\mu_\phi]$ is analyzed as the tilted angle around the $c$ axis, $\phi$, increases from $0$ to $20$ degrees.

\begin{figure}[!ht]
    \includegraphics[width=1\linewidth]{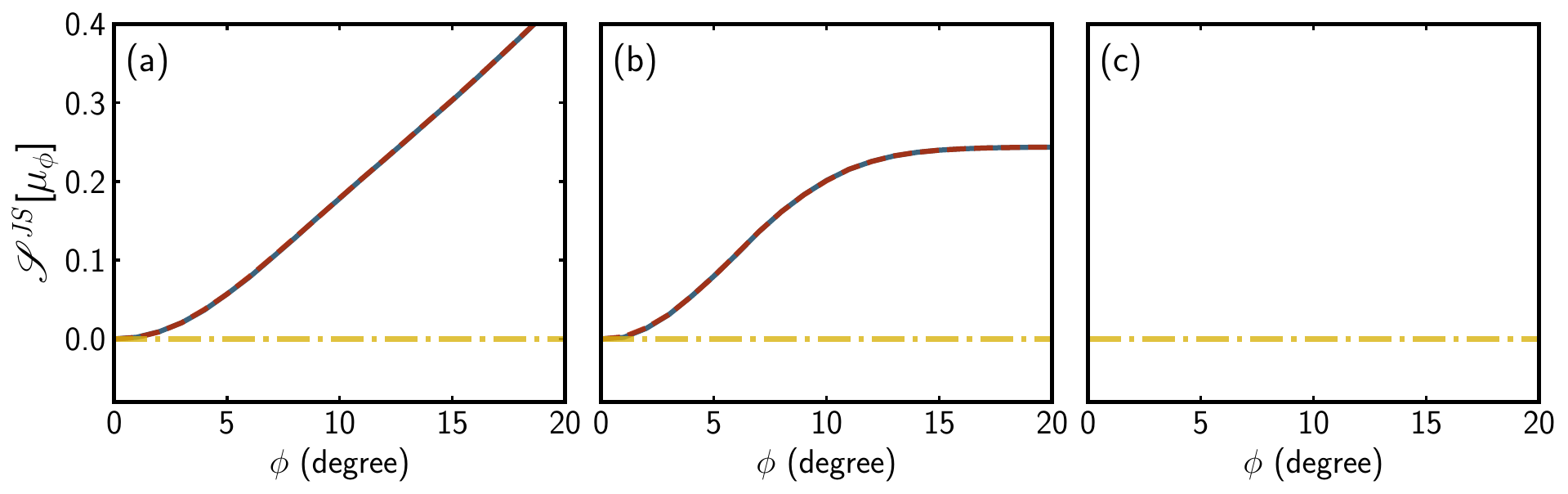}
\caption{The SBM $\sbm^{JS}[\mu_\phi]$ of a perovskite with in-phase octahedral rotations about an axis along the $c$ axis. 
(a) $\sbm^{JS}_{C_4}[\mu_\phi]$ is plotted for the four-fold rotation operator whose axis passes through the center Ti cation and points towards the $a$ axis (blue curve), the $b$ axis (red curve), and the $c$ axis (yellow curve). 
(b) $\sbm^{JS}_{\sigma_h}[\mu_\phi]$ is plotted for the reflection operator $\sigma_h$, whose mirror plane passes through the center Ti cation and is perpendicular to the $a$ axis (blue curve), the $b$ axis (red curve), and the $c$ axis (yellow curve). 
(c) $\sbm^{JS}_{i}[\mu_\phi]$ is plotted for the inversion operator $i$, whose fixed point is the central Ti cation. 
The figure is plotted as a function of the rotation angle $\phi$ in degree.}
\label{fig;SBM_perov}
\end{figure}

First, we investigate the SBM $\sbm^{JS}_{C_4}[\mu_\phi]$ associated with the four-fold rotation symmetry, and the results are illustrated in Fig.~\ref{fig;SBM_perov}(a). The operator's axis intersects the central Ti cation and is oriented along the $a$ axis (represented by the blue curve), $b$ axis (red curve), and $c$ axis (yellow curve). As the octahedral tilt increases, the $\sbm^{JS}_{C_4}[\mu_\phi]$ for $C_4$ rotation along the $c$ axis remains constant, as evidenced by the unvarying zero value depicted by the yellow line in Figure~\ref{fig;SBM_perov}(a). 
Conversely, the transition from the space group $Pm\bar{3}m$ (No.~221) to $P4/mbm$ (No.~127) results in the disruption of $C_4$ symmetry along both the $a$ and $b$ axes. By symmetry, $\sbm^{JS}_{C_4}[\mu_\phi]$ of $C_4$ along these two axes exhibit identical patterns of SB, as shown in the figure, where the $\sbm^{JS}_{C_4}[\mu_\phi]$ for $C_4$ rotation along the $a$ axis (blue curve) and the $b$ axis (red curve) display congruent monotonically increasing trajectories.

\paragraph{SBM of the reflection operator, $\sigma_h$}
The reflection plane intersects the central Ti cation, with its normal vector oriented towards the $a$ axis (illustrated by the blue curve), $b$ axis (red curve), and $c$ axis (yellow curve), as depicted in Figure~\ref{fig;SBM_perov}(b). 
Since there are no octahedral tilts around the $a$ and $b$ axes, reflection symmetry across the plane perpendicular to the $c$ axis is preserved. 
This is consistent with the yellow curve remaining at zero independent of \(\phi\), indicating the preservation of this symmetry under this distortion. 
Conversely, \(\sigma_h\) symmetry is disrupted over planes perpendicular to both the \(a\) and \(b\) axes, as reflected in the overlapping red and blue curves in Figure~\ref{fig;SBM_perov}(b). 
Notably, \(\sbm^{JS}_{\sigma_h}[\mu_\phi]\) converges to approximately $0.24$ beyond \(\phi=13\). Prior to \(\phi=10\), \(\sbm^{JS}_{\sigma_h}[\mu_\phi]\) exhibits a more rapid increase compared to \(\sbm^{JS}_{C_4}[\mu_\phi]\) and is strictly larger. 
However, post \(\phi=10\), \(\sbm^{JS}_{C_4}[\mu_\phi]\) accelerates in growth, while \(\sbm^{JS}_{\sigma_h}[\mu_\phi]\) approaches convergence, thus making \(\sbm^{JS}_{C_4}[\mu_\phi]\) greater than \(\sbm^{JS}_{\sigma_h}[\mu_\phi]\) for larger rotations. 
Consequently, the SBM analysis suggests that for smaller values of \(\phi\), the octahedral tilts \(a^0a^0c^+\) predominantly disrupt the reflection over planes perpendicular to the \(a\) and \(b\) axes, compared to rotation around these axes, whereas for larger \(\phi\) values, the inverse is observed.

\paragraph{SBM of the inversion operator, $i$}
We finally consider the inversion operator, $i$, defined with its fixed point at the central Ti cation. Since the octahedral tilts do not disrupt this inversion symmetry, $\sbm^{JS}_{i}[\mu_\phi]$ maintains a value of zero for all $\phi$ as expected, illustrated by the constant line in Figure~\ref{fig;SBM_perov}(c).

\subsection{Extensions and potential applications}

In the numerical tests presented above, we constructed the density $\mu$ using the Independent Atom Model (IAM), where each atom is represented by a Gaussian distribution derived from the average crystallographic ADPs (\uiso). This approach effectively captures the symmetry breaking of the
\emph{crystallographically averaged} structure. However, it does not account for correlated disorder.

Two-body correlations are experimentally accessible through the atomic pair distribution function (PDF) obtained from total-scattering experiments~\cite{egami;b;utbp12}. Accounting for such correlations could lead to different effective symmetry-breaking behavior, in which the symmetry breaking measure (SBM) returns different values depending on the range of interatomic distances over which it is evaluated. This situation arises, for example, in materials containing
statistically disordered, fluctuating, locally symmetry-broken domains~\cite{bozinLocalSnDipolarCharacter2023}, where the local, low-$r$ symmetry may be
lower than the domain-averaged symmetry observed at higher $r$.

The SBM is also affected by correlated atomic motions at short distances~\cite{jeongLatticeDynamicsCorrelated2003, desgrangesTemperaturedependentAnisotropyBond2023}. Due to such correlations (e.g., rigid-bond behavior), the distribution of instantaneous interatomic distances is often significantly narrower than that predicted by the convolution of uncorrelated thermal ellipsoids. This leads to smaller effective static displacements and, consequently, to larger changes in the SBM at small interatomic distances.

The JS-SBM framework presented here is not limited to the analytic IAM representation. Since the Jensen-Shannon divergence is defined for general probability densities, the method can be extended to incorporate correlated motion and dynamic effects. While the PDF is a two-point correlation function rather than a probability density, it is a continuous function in $\mathbb{R}^3$, and a valid
probability distribution can be constructed through appropriate normalization. For example, we propose
\begin{equation}
    P_g(r) = \frac{4\pi r^2 \rho(r)}{\int_0^{r_{\max}} 4\pi r^2 \rho(r)\,\mathrm{d}r},
\end{equation}
where $r_{\max}$ is the maximum distance considered, and $\rho(r)$ is the atomic-pair density defined as the number of atoms found in a spherical shell of thickness $\mathrm{d}r$ at a distance $r$ from a reference atom, averaged over all atoms in the sample~\cite{farrowRelationshipAtomicPair2009, egami;b;utbp12}. This quantity can be computed directly from structural models and, notably, from the experimentally accessible $G(r)$ function via
\begin{equation}
    \rho(r) = \frac{G(r)}{4\pi r} - \gamma_0(r)\rho_0.
\end{equation}
Here $\rho_0$ is the average atomic density, and $\gamma_0(r)$ is the characteristic function of the sample. It takes a value of unity for an infinite crystal and, for finite particles, represents the orientationally averaged autocorrelation of the particle shape function, giving the probability that two points separated by a distance $r$ both lie within the particle.

Although we have not carried out a detailed analysis of the SBM applied to this distribution, such a calculation would be straightforward to perform, for example, on ensembles of instantaneous configurations generated using Molecular Dynamics (MD) or Reverse Monte Carlo (RMC) simulations. In this context, the SBM would quantify \emph{instantaneous symmetry breaking}, capturing local distortions that may average out in time-averaged structures but remain visible in PDF measurements. Moreover, since $G(r)$ is directly accessible experimentally, this PDF-based SBM could be computed without first solving the structure, for example in a temperature-dependent study across a structural phase transition.

The ability to define SBMs for different probability distributions also suggests that, more generally, a family of symmetry-breaking measures may be constructed for material systems, potentially revealing insights that are not accessible through a single representation. For example, Fang \emph{et al.}~\cite{fangQuantificationSymmetry2016} introduced a continuous symmetry measure that quantifies changes in a system's Hamiltonian, rather than the downstream effects on the IAM or the PDF.

This flexibility suggests that the SBM could serve as a valuable tool within modeling frameworks such as \textsc{RMCProfile}~\cite{tuckerRMCProfileReverseMonte2007} or \textsc{PDFgui}~\cite{farro;jpcm07, yangXPDFsuiteEndtoendSoftware2015}. By incorporating the SBM as a continuous order parameter or as a regularization term in the loss function, researchers could quantitatively monitor specific symmetry violations during refinement, enabling a more controlled exploration of disorder and local symmetry breaking.

\section{Availability of code}
\label{sec:code}

The software code used to produce the results shown in this paper, including the implementation of the SBM algorithms, is publicly accessible at the GitHub repository: \url{https://github.com/lanikaling/SymmetryBreakingMeasure}. This repository contains detailed instructions for calculating the SBM for general finite clusters. Users can manually define their own finite clusters or import unit cell data directly from a CIF file. The code supports applying distortions and calculating the SBM for various symmetry operations, such as rotation, reflection, and inversion. Furthermore, we plan to eventually integrate this functionality into the DiffPy organization as a dedicated package (\url{https://github.com/diffpy/diffpy.sbm}) to provide a standardized open-source distribution in the future.

\section{Conclusion}
In this paper, we introduce a continuous SBM using the Jensen-Shannon divergence to analyze structural transformations and distortions. 
It is designed for studying SB in finite clusters where SB is continuous and there is value in quantifying it. 
In infinite crystals, a symmetry is either present or broken, and it cannot be slightly broken. The continuous SBM, on the contrary, provides insights into local SB that would otherwise be obscured in the analysis of the infinite crystal. 

Typically, for a finite cluster undergoing collective displacements of multiple atoms, traditional methods can only determine which symmetries are preserved or disrupted, without providing a comparative assessment of the extent of SB across different symmetry operations. In contrast, our SBM has proven to be a robust tool that can offer a quantitative framework for evaluating and comparing the degree of SB caused by various distortions. This approach not only identifies which specific symmetry operation experiences greater violation due to the distortion but also enriches the descriptive language available for quantitatively discussing structural distortions.

In recent years, with the growing application of machine learning (ML) in structural science~\cite{billingeMachineLearningCrystallography2024}, quantifying and implementing SB has become essential for identifying and discovering symmetries. In supervised tasks like symmetry discovery, this can involve either determining a numerical label that describes the symmetries of the input dataset or incorporating SB directly into the loss function. For example, Forestano \emph{et~al.}~\cite{forestanoDeepLearningSymmetries2023} treat symmetry labeling as a black-box regression problem, while Liu \emph{et~al.}~\cite{liuMachineLearningHidden2022} incorporate symmetry into the loss function by quantifying SB as the violation of certain partial differential equations. 
Our SBM is well-suited for quantifying SB in these cases, as it is computationally efficient, applies to any finite cluster, and accounts for not only particle positions but also particle type and thermal vibrations. 
Additionally, in ML models for material prediction, SBM can serve as a bounded regularization term to enforce preferred symmetries.

\section*{Acknowledgment}
Work in the Billinge group was supported by U.S. Department of Energy, Office of Science, Office of Basic Energy Sciences (DOE-BES) under contract No.~DE-SC0024141. 
The work of Qiang Du is supported in part by No.~DE-SC0022317 and No.~DE-SC0025347 from the Department of Energy.

\appendix \section{Appendix A: Derivation of Symmetry Breaking Measure for a Single-atom System}\label{app:single_atom_proof}
Let $\mu(\mathbf x)=\mathcal N(\mathbf 0, U\cdot \bm{I}_3)$ and $(T_{\bm{d}})_\#\mu(\mathbf x)=\mathcal N(\mathbf d, U\cdot \bm{I}_3)$. Writing the densities,
\begin{equation}
\mu(\mathbf x)=(2\pi U)^{-3/2}\exp\!\Big(-\tfrac{1}{2U}\|\mathbf x\|^2\Big),\quad
(T_{\bm{d}})_\#\mu(\mathbf x)=(2\pi U)^{-3/2}\exp\!\Big(-\tfrac{1}{2U}\|\mathbf x-\mathbf d\|^2\Big).
\end{equation}
Since the normalizing constants coincide, the log-ratio simplifies to
\begin{equation}
\log\frac{\mu(\mathbf x)}{(T_{\bm{d}})_\#\mu(\mathbf x)}
= -\frac{1}{2U}\|\mathbf x\|^2+\frac{1}{2U}\|\mathbf x-\mathbf d\|^2
= \frac{1}{2U}\!\left(\|\mathbf x-\mathbf d\|^2-\|\mathbf x\|^2\right)
= \frac{1}{2U}\left(\|\mathbf d\|^2-2\,\mathbf x^\top\mathbf d\right).
\end{equation}
Therefore,
$$D_{\mathrm{KL}}(\mu\|(T_{\bm{d}})_\#\mu)=\mathbb E_{\mu}\!\left[\log\frac{\mu(\mathbf x)}{(T_{\bm{d}})_\#\mu(\mathbf x)}\right]
= \frac{1}{2U}\left(\|\mathbf d\|^2-2\,\mathbf d^\top \mathbb E_{\mu}[\mathbf x]\right).$$
Because $\mu$ is centered and spherically symmetric, $\mathbb E_{\mu}[\mathbf x]=\mathbf 0$. Thus, the linear cross term vanishes and we obtain the expression
$$D_{\mathrm{KL}}(\mu\|(T_{\bm{d}})_\#\mu)=\frac{\|\mathbf d\|^2}{2U} = \frac{d^2}{2U},$$
where $d=\|\bm{d}\|_{L_2}$ is the displacement distance. 

\bibliographystyle{apsrev4-2}
\bibliography{symmetry_breaking_measure}

\end{document}